\documentclass[12pt]{article}
\makeatletter
\@addtoreset{equation}{section}
\makeatother

\begin{document}
\begin{center}
{\Large \bf Unboundable Spacetimes with Metric Singularities
and Matching Metrics and Geodesics:\\[0.6cm]

A Black-White Hole and a Big Crunch-Bang} \\[1.5cm]
{\bf Vladimir S.~MASHKEVICH}\footnote {E-mail:
Vladimir\_Mashkevich@qc.edu}  \\[1.4cm]
{\it Queens College\\
The City University of New York\\
65-30 Kissena Boulevard\\
Flushing, New York 11367-1519}\\[1.4cm]
\vskip 1cm

{\large \bf Abstract}
\end{center}

Singularity theorems of general relativity utilize the notion
of causal geodesic incompleteness as a criterion of the
presence of a spacetime singularity. The incompleteness  of
a causal curve implies the end and/or beginning of the
existence of a particle, which are events. In the commonly
accepted approach, singularities are not incorporated into
spacetime. Thus spacetime turns out to be event-incomplete.
With creation from nothing, singularities are sources
of lawlessness. A straightforward way around these conceptual
problems consists in including metric singularities in
spacetime and then matching metrics and causal geodesics at
the singularities. To this end, a spacetime manifold is
assumed to be unboundable,
so that singularities may only be interior.
The matching of the geodesics is
achieved through weakening conditions for their smoothness.
This approach is applied to a black-white hole and a big
crunch-bang.

\newpage

\section*{Introduction}

Spacetime singularities are inherent in general relativity,
or more specifically in gravitational collapse and cosmology.
The analysis of spacetimes with singularities is one of the
most principal and difficult problems in general relativity.
Singularity theorems of general relativity utilize the notion
of causal geodesic incompleteness as a criterion for the
presence of a singularity. (A comprehensive presentation and
discussion is given in [1].) The incompleteness of a causal,
i.e., timelike or null curve implies physically the end and/or
beginning of the existence of a particle, which are
undeniably events. In the commonly accepted approach,
singularities are not incorporated into a spacetime manifold.
Thus spacetime turns out to be event-incomplete, i.e.,
does not include all events.

Furthermore, the beginning and end of the existence of a free
particle means that there is creation from nothing and
extinction into nothing. Those phenomena are in conflict with
conservation laws and appear physically pathological. Maybe
it is possible to put up with extinction into nothing,
arguing that nature is so structured. At least extinction follows
a clear-cut law: Arriving at a singularity results in
extinction. With creation from nothing, the situation is
much worse. In this role, (naked) singularities are sources of
lawlessness. All sorts of nasty things---green slime, Japanese
horror movie monsters, etc.---may emerge helter-skelter from a
singularity [1]. To get rid of that nightmare, Penrose proposed
the cosmic censorship hypothesis. But cosmic censorship may be
legislated only by a fiat, it does not follow from known physical
laws.

Relying on the aforesaid, we state that there exists the
conceptual
problem of singularities, which should be dealt with. We shall
restrict our consideration to metric singularities and
their related
causal geodesic incompleteness. A straightforward
approach consists in including the singularities in spacetime
and making them to be interior, not exterior ones, and then
matching metrics and causal geodesics at the singularities.
To this end, a spacetime manifold is assumed to be unboundable,
i.e., to be a manifold without boundary and not to be a manifold
with boundary, the latter being removed. Such a manifold is
realized as a closed submanifold of a Euclidean space. The
matching of metrics is based on matching tangents to curves
through a singularity. The matching of causal geodesics is
achieved through weakening conditions for their smoothness. The
main remaining condition is that of cornerlessness.

The next step is made by considering a product spacetime. This
allows one to formulate the conditions of smoothness
in terms of time
dependence of the geodesics. In particular, the examination of
acceleration is helpful in matching geodesics.

The approach outlined above is applied to a black-white hole
and a big crunch-bang. Synchronous coordinates are utilized.
Radial geodesics are investigated.

The main results for black-white holes are as follows. There are
two types of metric singularities: singular three-dimensional
interfaces between black and white regions and singular
threefolds within white regions. On the former surfaces particles
reflect from hypersurfaces of constant radius, on the latter
particles traverse through those.

For a big crunch-bang there is a metric-singular
three-dimensional interface between contracting and expanding
regions. On this surface particles traverse through
hypersurfaces of constant radius.

\section{A conventional concept of a singular spacetime
and event-incompleteness of the latter}

Our ultimate goal is to surmount causal geodesic incompleteness.
The first step is to include singular points in a spacetime
manifold. In a conventional treatment, such an inclusion is
believed to be impossible. The main argument is that physical
laws are violated at singular points. But if a set of the
singular points is a hypersurface, the violation may be overcome
through matching results of the laws at points of the
hypersurface. On the other hand, excluding singular points from
a spacetime manifold implies excluding corresponding events,
which means event-incompleteness of spacetime. But spacetime is
by definition the set of all events. Thus the inclusion of
singularities in the spacetime manifold seems to be justified.

\section{An unboundable manifold and a closed submanifold
as its realization}

Singularities not only have to be included in the spacetime
manifold $M$, but also should not be localized on a boundary
of $M$; for otherwise causal curves cannot be extended. The
problem of boundary may be posed more generally. A boundary
of $M$ or a possibility of attaching one to $M$ give rise to
curve incompleteness. To prevent this we arrive at the notion
of an unboundable manifold, i.e., a manifold without boundary
to which no boundary can be attached:
\begin{equation}
M=({\rm manifold\: without\: boundary})\;{\rm {\rm and}}\;M\ne(
{\rm manifold\: with\:
boundary})-({\rm boundary})
\label{2.1}
\end{equation}

An unboundable manifold may be realized as a closed subset of
an Euclidean space by the Whitney theorem [2]: A smooth manifold
$M^{n}$ can be embedded as a submanifold, and closed subset, of
$R^{2n+1}$.

So we posit spacetime to be an unboundable manifold.

\section{Interior singularities and matching metrics
and\protect \\
geodesics}

{\it Interior singularities\/}

\noindent Now that singularities may only be interior,
the problem of
overcoming causal geodesic incompleteness amounts not to
extending a manifold and geodesics but rather to matching metrics
and geodesics on opposite sides of a metric singularity
hypersurface.
\par\medskip
\noindent {\it Matching metrics\/}

\noindent First consider metrics. Let $p\in M$ be a point
on a metric
singularity hypersurface and $\gamma(u)$ be a $C^{\infty}$
curve through this point such that
\begin{equation}
\gamma(u_{\rm sing})=p,\quad \dot \gamma(u_{\rm sing})\ne 0
\label{3.1}
\end{equation}
and $\dot \gamma(u_{\rm sing})$ is not tangent to the
hypersurface.
Matching conditions are: for all $\gamma$ specified
\begin{equation}
\int_{u_{\rm sing}-\delta}^{u_{\rm sing}+\delta}|g_{\gamma(u)}
(\dot \gamma(u),\dot \gamma(u))|^{1/2}du<\infty
\label{3.2}
\end{equation}
and
\begin{equation}
  \lim_{\delta\to 0}\frac{g_{\gamma(u_{\rm sing}+\delta)}
(\dot \gamma(u_{\rm sing}+\delta),\dot \gamma(u_{\rm sing}+
\delta))}
{g_{\gamma(u_{\rm sing}-\delta)}(\dot \gamma(u_{\rm sing}-
\delta),
\dot \gamma(u_{\rm sing}-\delta))}=1
\label{3.3}
\end{equation}
\par\medskip
\noindent{\it Matching geodesics\/}

\noindent Now turn to geodesics. Geodesic equations are
of the form
\begin{equation}
\frac{dK^{\mu}}{du}+\Gamma^{\mu}_{\nu\rho}K^{\nu}K^{\rho}=
0,\quad K^{\mu}=\frac{dx^{\mu}}{du}
\label{3.4}
\end{equation}
We put
\begin{equation}
du=\frac{1}{m}ds
\label{3.5}
\end{equation}
for a particle of a mass $m\ne 0$. Then
\begin{equation}
K^{\mu}K_{\mu}=m^{2}
\label{3.6}
\end{equation}
holds for both $m\ne 0$ and $m=0$.

Let $x_{\rm sing}^{\mu}$ be coordinates of $p$ and
$x^{\mu}(u)$ be coordinates of two causal geodesics $\gamma(u)$,
$u<u_{\rm sing}$ or $u>u_{\rm sing}$, on opposite
sides of the singular
hypersurface. Matching conditions are:
\begin{equation}
\lim_{\delta\to 0}\gamma(u_{\rm sing}+\delta)=
\lim_{\delta\to 0}\gamma(u_{\rm sing}-\delta)=p
\label{3.7}
\end{equation}
and\begin{equation}
\lim_{\delta\to 0}\frac{(d^{k}x^{\mu}/du^{k})_{u_{\rm sing}
+\delta}}
{(d^{k}x^{\mu}/du^{k})_{u_{\rm sing}-\delta}}=1,\;k=
1,2,\ldots,k_{{\rm maximal}}
\label{3.8}
\end{equation}
\begin{equation}
\lim_{\delta\to 0}\left( \frac{d^{k}x^{\mu}}{du^{k}} \right)
_{u_{\rm sing}\pm \delta}= +\infty\quad {\rm or}\quad -\infty,
\quad k=k_{\rm maximal}
\label{3.9}
\end{equation}
being admissible. Such a curve may be termed a $\bar{C}^{k}$
curve (extended $C^{k}$ curve).

\section{A product spacetime}

A product spacetime manifold is a typical and the most important
instantiation of the notion of an unboundable spacetime
manifold. The manifold $M=M^{4}$ is a direct product of two
manifolds:
\begin{equation}
M=T\times S,\quad M\ni p=(t,s),\quad t\in T,\quad
-\infty<t<\infty,\quad s\in S
\label{4.1}
\end{equation}
The one-dimensional unboundable manifold $T$ is time, the
three-dimensional manifold $S$, which should be unboundable, is
a space. By (\ref{4.1}) the tangent space at a point
$p\in M$ is\begin{equation}
M_{p}=T_{p}\oplus S_{p}
\label{4.2}
\end{equation}
Assuming that
\begin{equation}
T_{p}\perp S_{p}
\label{4.3}
\end{equation}
it follows for the metric tensor that
\begin{equation}
g=g_{T}+g_{S}
\label{4.4}
\end{equation}
Furthermore,\begin{equation}
g=dt\otimes dt-h_{t}
\label{4.5}
\end{equation}
where $h_{t}$ is a Riemannian metric tensor on $S$ depending
on $t$. A relevant coordinate representation is
\begin{equation}
ds^{2}=dt^{2}-dh^{2}_{t},\qquad dh^{2}_{t}=h_{tij}^{2}
dx^{i}dx^{j}
\label{4.6}
\end{equation}
$(t=x^{0},x^{1},x^{2},x^{3})$ are synchronous coordinates.

Now the problem of metric singularities amounts to that for
the metric $h$. The problem is thereby greatly simplified
since $h$ is a Riemannian metric, for which there is a fully
satisfactory notion of the location of singularities [3].

In the case of the metric (\ref{4.6}), equation (\ref{3.6})
reduces to
\begin{equation}
(K^{0})^{2}-h_{ij}K^{i}K^{j}=m^{2}
\label{4.7}
\end{equation}
or
\begin{equation}
\omega^{2}-K_{i}K^{j}=m^{2},\qquad \omega=E=K^{0}
\label{4.8}
\end{equation}

\section{Matching geodesics: Refinement}

Let us return to matching causal geodesics and employ the
universal time $t$ as a parameter. So a geodesic is given by
functions $x^{i}(t)$. Let the geodesic pass through a singular
point $p$. Choose coordinates so that $p=(0,0,0,0)$ and
$x^{2}(t)=x^{3}(t)=0$ along the geodesic for $t\approx 0$. Then
the geodesic is described by a function $x(t)\equiv x^{1}(t)$.
The geodesic is at least $\bar{C}^{1}$, so that we have for
$t\approx 0$
\begin{equation}
|x(t)|\approx |x(-t)|
\label{5.1}
\end{equation}
which implies
\begin{equation}
|\dot x(t)|\approx|\dot x(-t)|,\qquad |\ddot x(t)|\approx
|\ddot x(-t)|
\label{5.2}
\end{equation}
Furthermore, in view of singularity
\begin{equation}
\lim_{t\to 0}|\ddot x(t)|=\infty
\label{5.3}
\end{equation}

There are two possibilities:
\begin{equation}
(1)\;\dot x(0)\ne 0,\;\lim_{t\to 0}|\dot x(t)|=\infty,\;
\dot x(t)\approx\dot x(-t),\;x(t)\approx -x(-t),\;
\ddot x(t)\approx -\ddot x(t),\;{\rm sgn}\ddot x(t)=-
{\rm sgn}x(t)
\label{5.4}
\end{equation}
this is the case of attraction, the particle transverses the
point $x=0$;
\begin{equation}
(2)\quad \dot x(0)=0,\qquad \dot x(t)\approx -\dot x(-t),
\qquad x(t)\approx
x(-t),\qquad \ddot x(t)\approx\ddot x(-t),\qquad {\rm sgn}\ddot x(t)=
{\rm sgn}x(t)
\label{5.5}
\end{equation}
this is the case of repulsion, the particle reflects from the
point $x=0$.

\section{A black-white hole}

{\it Spacetime manifold\/}

\noindent Spacetime manifold of a spherically symmetric
black-white hole
is a product manifold (\ref{4.1}), the space $S$ being
three-dimensional Euclidean space $R^{3}$:
\begin{equation}
M=T\times R^{3}
\label{6.1}
\end{equation}
\par\medskip
\noindent{\it Synchronous coordinates\/}

\noindent Synchronous coordinates are $t$ and spatial
coordinates for
$R^{3}$. In view of spherical symmetry, the spherical
coordinates $(R,\theta,\phi)$ are appropriate with $R=0$
in the center of the star. Let the surface of the star
correspond to $R=a$ ($a={\rm const}$ in synchronous coordinates).
We are interested in the region outside the star, $R>a$.
\par\medskip
\noindent{\it Metric\/}

\noindent The metric $dh^{2}$ is given as follows [4]. For
$R>a$
\begin{equation}
dh^{2}=\frac{[\partial _{R}r(t,R)]^{2}}{1+f(R)}dR^{2}+
r^{2}(t,R)(d\theta^{2}+\sin^{2}\theta d\phi^{2})
\label{6.2}
\end{equation}
\begin{equation}
r=-\frac{r_{g}}{f(R)}\frac{1+\cos\eta}{2},\quad
t_{0}(R)
-t=\frac{r_{g}}{[-f(R)]^{3/2}}\frac{\pi-\eta-\sin\eta}{2}
\label{6.3}
\end{equation}
where $t_{0}(R)$ is an arbitrary function, $f(R)$
is an arbitrary function meeting the condition
\begin{equation}
-1<f(R)<0
\label{6.4}
\end{equation}
and $r_{g}$ is the Schwarzschild radius. We put
\begin{equation}
t_{0}(R)=\frac{r_{g}}{[-f(R)]^{3/2}}\frac{\pi}{2}
\label{6.5}
\end{equation}
so that
\begin{equation}
r=-\frac{r_{g}}{f(R)}\frac{1+\cos\eta}{2},\quad
t=\frac{r_{g}}{[-f(R)]^{3/2}}\frac{\eta+\sin\eta}{2}
\label{6.6}
\end{equation}
Now we choose
\begin{equation}
f(R)=-\frac{r_{g}}{R}
\label{6.7}
\end{equation}
In view of
\begin{equation}
R>a>r_{g}
\label{6.8}
\end{equation}
the condition (\ref{6.4}) holds. We obtain
\begin{equation}
r=R\frac{1+\cos\eta}{2},\quad t=\frac{R^{3/2}}{r_{g}^{1/2}}
\frac{\eta+\sin \eta}{2}
\label{6.9}
\end{equation}
\begin{equation}
dh^{2}=\frac{(\partial _{R}r)^{2}}{1-r_{g}/R}dR^{2}+
r^{2}(d\theta^{2}+\sin^{2}\theta d\phi^{2})
\label{6.10}
\end{equation}
For what follows, it is convenient to introduce quantities
\begin{equation}
\chi=\frac{1+\cos\eta}{2},\qquad\xi=\frac{\eta+\sin\eta}{2}
\label{6.11}
\end{equation}
so that
\begin{equation}
\chi=\chi(\xi),\qquad \xi=\frac{r_{g}^{1/2}t}{R^{3/2}},\qquad
r(t,R)=R\chi\left( \frac{r_{g}^{1/2}t}{R^{3/2}} \right)
\label{6.12}
\end{equation}
the metric being given by (\ref{6.10}).
\par\medskip
\noindent{\it Metric discontinuity on the surface of
the star\/}

\noindent Note that in synchronous coordinates, discontinuity
of the matter density on the surface of the star results in
a metric discontinuity as well, which is easily seen from [4,5].
\par\medskip
\noindent{\it Metric singularities\/}

\noindent It follows from (\ref{6.11}), (\ref{6.12}) that
\begin{equation}
\partial _{R}r=\chi-\frac{3}{2}\xi\frac{d\chi}{d\xi}
\label{6.13}
\end{equation}
\begin{equation}
\frac{d\chi}{d\xi}=-\frac{\sin\eta}{1+\cos\eta},\qquad
\left( \frac{d\chi}{d\xi} \right)^{2}=\frac{1}{\chi}-1
\label{6.14}
\end{equation}
\begin{equation}
\partial _{R}r=\frac{1+\cos\eta}{2}+\frac{3}{2}\frac{\sin\eta}
{1+\cos\eta}\frac{\eta+\sin\eta}{2}
\label{6.15}
\end{equation}

There are two types of metric singularities:
\begin{equation}
r=0
\label{6.16}
\end{equation}
which implies
\begin{equation}
\chi=0,\qquad \left( \frac{d\chi}{d\xi} \right)^{2}=\infty,
\qquad (\partial _{R}r)^{2}=\infty
\label{6.17}
\end{equation}
and
\begin{equation}
\partial _{R}r=0
\label{6.18}
\end{equation}
which implies
\begin{equation}
\frac{3}{2}\eta\sin\eta+\frac{3}{2}\sin^{2}\eta+
(1+\cos\eta)^{2}=0
\label{6.19}
\end{equation}

Let
\begin{equation}
t\ge 0
\label{6.20}
\end{equation}
(for example, $t=0$ corresponds to the beginning of the collapse).
Then
\begin{equation}
\xi\ge 0,\qquad \eta\ge 0
\label{6.21}
\end{equation}

For the singularities (\ref{6.16}) we obtain
\begin{equation}
\eta=(2n+1)\pi,\;n=0,1,2,\ldots,\qquad \xi=(n+1/2)\pi\equiv
\xi_{n+1/2}
\label{6.22}
\end{equation}
and the equation for the singular hypersurfaces is of the form
\begin{equation}
\frac{r_{g}^{1/2}t}{R^{3/2}}=\xi_{n+1/2},\;n=0,1,2,\ldots
\label{6.23}
\end{equation}

For the singularities (\ref{6.18}) we have $\sin\eta<0$ so that
\begin{equation}
\eta=(n+1)2\pi-\beta,\quad 0<\beta<\pi,\quad n=0,1,2,\ldots
\label{6.24}
\end{equation}
The equation for $\beta$ is of the form
\begin{equation}
[(n+1)3\pi-\frac{3}{2}(\beta+\sin\beta)]\sin\beta-
(1+\cos\beta)^{2}=0,\quad 0<\beta<\pi,\quad n=0,1,2,\ldots,
\quad \beta=\beta_{n+1}
\label{6.25}
\end{equation}
Corresponding values of $\xi$ are
\begin{equation}
\xi_{\beta,n+1}=(n+1)\pi-\frac{1}{2}(\beta_{n+1}+
\sin\beta_{n+1}),\quad n=0,1,2,\ldots
\label{6.26}
\end{equation}
and the equation for the singular hypersurfaces is of the form
\begin{equation}
\frac{r_{g}^{1/2}t}{R^{3/2}}=\xi_{\beta,n+1},\quad n=0,1,2,\ldots
\label{6.27}
\end{equation}
\par\medskip
\noindent{\it Matching metrics\/}

\noindent The only parameter that appears in the metric is
the Schwarzschild radius $r_{g}$. Therefore matching metrics at
singularities amounts to taking the same value of the star mass
for all the regions of the spacetime.
\par\medskip
\noindent{\it Black and white regions\/}

\noindent According to (\ref{6.11}) $\chi$ changes between
values 1 and 0, \begin{equation}
\chi(\xi_{n})=1,\quad \chi(\xi_{n+1/2})=0,\quad \xi_{n}=n\pi,
\quad \xi_{n+1/2}=(n+1/2)\pi,\quad n=0,1,2,\ldots
\label{6.28}
\end{equation}
and $\chi(\xi)$ decreases for $\xi_{n}<\xi<\xi_{n+1/2}$ and
increases for $\xi_{n+1/2}<\xi<\xi_{n+1}$. Thus, in view of
(\ref{6.12}), the regions of black and white hole are,
respectively,
\begin{equation}
\frac{r_{g}^{1/2}t}{R^{3/2}}=\xi,\quad {\rm black}
:\;\xi_{n}<\xi<
\xi_{n+1/2},\quad {\rm white}:\;\xi_{n+1/2}
<\xi<\xi_{n+1},\quad
n=0,1,2,\ldots
\label{g6.29}
\end{equation}
The singularities (\ref{6.16}) correspond to passages
black$\to$white (black-white singularities), the singularities
(\ref{6.18}) lie in white regions (white singularities).
\par\medskip
\noindent{\it Geodesic equations\/}

\noindent We shall consider radial geodesics:
\begin{equation}
(K^{i})=(K^{R},0,0),\qquad \omega^{2}-m^{2}=h_{RR}(K^{R})^{2}
\label{6.30}
\end{equation}
where by (\ref{6.10})
\begin{equation}
h_{RR}=\frac{(\partial _{R}r)^{2}}{1-r_{g}/R}
\label{6.31}
\end{equation}
We have
\begin{equation}
\left( \frac{K^{R}}{K^{0}} \right)^{2}=\frac{\omega^{2}-
m^{2}}{\omega^{2}}\frac{1}{h_{RR}}
\label{6.32}
\end{equation}

Equations (\ref{3.4}) boil down to [5]
\begin{equation}
\frac{dK^{R}}{du}+\Gamma^{R}_{RR}(K^{R})^{2}+2\Gamma_{0R}^{R}
K^{0}K^{R}=0
\label{6.33}
\end{equation}
\begin{equation}
\frac{dK^{0}}{du}=\Gamma_{RR}^{0}(K^{R})^{2}=0
\label{6.34}
\end{equation}
with
\begin{equation}
\Gamma_{RR}^{R}=\frac{1}{2}\frac{\partial _{R}h_{RR}}
{h_{RR}},\quad \Gamma_{0R}^{R}=\frac{1}{2}\frac{\partial
_{t}h_{RR}}{h_{RR}},\quad \Gamma_{RR}^{0}=\frac{1}{2}
\partial _{t}h_{RR}
\label{6.35}
\end{equation}

A trivial, familiar solution is
\begin{equation}
K^{R}=0,\quad R={\rm const},\qquad K^{0}=\omega=m\ne 0
\label{6.36}
\end{equation}
---there is no problem of matching for it.
\par\medskip
\noindent{\it Matching geodesics at a black-white
singularity\/}

\noindent In the vicinity of a black-white hole singularity
(\ref{6.16}), (\ref{6.17}) we make use of equation (\ref{6.34}).
With (\ref{6.35}), (\ref{6.32}), and (\ref{6.31}) we obtain
\begin{equation}
\frac{dK^{0}}{du}=\frac{\partial _{t}\partial _{R}r}{\partial
_{R}r}\frac{(K^{0})^{2}-m^{2}}{K^{0}}\frac{dt}{du}=0
\label{6.37}
\end{equation}
whence
\begin{equation}
\frac{d}{dt}\left[ (K^{0})^{2}-m^{2} \right]+2\frac
{\partial _{t}\partial _{R}r}{\partial _{R}r}
\left[ (K^{0})^{2}-m^{2} \right]=0
\label{6.38}
\end{equation}
{}From (\ref{6.32}) follows
\begin{equation}
\left( \frac{dR}{dt} \right)^{2}=\frac{(K^{0})^{2}-m^{2}}
{(K^{0})^{2}}\frac{1-r_{g}/R}{(\partial _{R}r)^{2}}
\label{6.39}
\end{equation}
We find from (\ref{6.12}), (\ref{6.11}) in the vicinity of
$r=0$
\begin{equation}
\partial _{R}r\approx -\left( \frac{3}{2} \right)^{2/3}
\xi_{n+1/2}\frac{1}{(\xi-\xi_{n+1/2})^{1/3}},\quad
\frac{\partial _{t}\partial _{R}r}{\partial _{R}r}\approx
-\frac{1}{3}\frac{\xi_{n+1/2}}{t_{s}}\frac{1}{\xi-\xi_{n+1/2}}
\label{6.40}
\end{equation}
where $t_{s}=t_{\rm singularity}$. By (\ref{6.39})
\begin{equation}
\frac{dR}{dt}\approx 0
\label{6.41}
\end{equation}
so that
\begin{equation}
\xi-\xi_{n+1/2}\approx \frac{\xi_{n+1/2}}{t_{s}}(t-t_{s})
\label{6.42}
\end{equation}
Thus we obtain from (\ref{6.38})
\begin{equation}
\frac{d}{dt}\left[ (K^{0})^{2}-m^{2} \right]-\frac{2}{3}
\frac{1}{t-t_{s}}\left[ (K^{0})^{2}-m^{2} \right]\approx 0
\label{6.43}
\end{equation}
{}from where
\begin{equation}
(K^{0})^{2}\approx m^{2}+A^{2}(t-t_{s})^{2/3}
\label{6.44}
\end{equation}
Now (\ref{6.39}) gives
\begin{equation}
\left( \frac{dR}{dt} \right)^{2}\approx b^{2}
\frac{A^{2}(t-t_{s})^{4/3}}{m^{2}+A^{2}(t-t_{s})^{2/3}},\quad
b^{2}=\left( \frac{2}{3} \right)^{4/3}\frac{R_{s}^{2}
(1-r_{g}/R_{s})}{r_{g}^{2/3}t_{s}^{2}}
\label{6.45}
\end{equation}
whence
\begin{equation}
\frac{dR}{dt}\approx |b|\frac{A(t-t_{s})^{2/3}}{\sqrt{
m^{2}+A^{2}(t-t_{s})^{2/3}}}=|b|
\left\{
\begin{array}{rcl}
\displaystyle\frac{A}{m}(t-t_{s})^{2/3},\;m\ne 0\\
\\
\displaystyle\frac{A}{|A|}|t-t_{s}|^{1/3},\;m=0\\
\end{array}
\right.
\equiv \tilde{A}|t-t_{s}|^{\beta},\quad 0<\beta<1
\label{6.46}
\end{equation}
We obtain
\begin{equation}
R-R_{s}\approx \tilde{A}\frac{1}{1+\beta}|t-t_{s}|^{1+\beta}
{\rm sgn}(t-t_{s}),\quad \frac{d^{2}R}{dt^{2}}\approx \tilde{A}
\beta\frac{1}{|t-t_{s}|^{1-\beta}}{\rm sgn}(t-t_{s})
\label{6.47}
\end{equation}
In fact, (\ref{6.46}), (\ref{6.47}) describe two solutions:
for $t<t_{s}$ and $t>t_{s}$, so that
\begin{equation}
\tilde{A}=\tilde{A}^{[{\rm sgn}(t-t_{s})]}
\label{6.48}
\end{equation}
Thus
\begin{equation}
\frac{d^{2}R}{dt^{2}}\approx \tilde{A}^{[{\rm sgn}(t-t_{s})]}
\beta\frac{1}{|t-t_{s}|^{1-\beta}}{\rm sgn}(t-t_{s})
\label{6.49}
\end{equation}
The quantities $R$ and $dR/dt$ are continuous. In order that
$R(t)$ be maximally smooth
\begin{equation}
\lim_{\delta \to 0}\frac{(d^{2}R/dt^{2})_{t_{s}+\delta}}
{(d^{2}R/dt^{2})_{t_{s}-\delta}}=1
\label{6.50}
\end{equation}
should hold, whence
\begin{equation}
-\tilde{A}^{[-1]}=\tilde{A}^{[+1]}\equiv \tilde{A},\quad
\tilde{A}^{[{\rm sgn}(t-t_{s})]}=\tilde{A}{\rm sgn}(t-t_{s})
\label{6.51}
\end{equation}
so that
\begin{equation}
R-R_{s}\approx \tilde{A}\frac{1}{1+\beta}|t-t_{s}|^{1+\beta},
\quad\frac{dR}{dt}\approx \tilde{A}|t-t_{s}|^{\beta}
{\rm sgn}(t-t_{s}),\quad \frac{d^{2}R}{dt^{2}}\approx
\tilde{A}\beta\frac{1}{|t-t_{s}|^{1-\beta}}
\label{6.52}
\end{equation}
We find
\begin{equation}
\frac{d^{2}R}{dt^{2}}\approx \frac{\beta}{(1+\beta)
^{(1-\beta)/(1+\beta)}}\tilde{A}^{2/(1+\beta)}
\frac{R-R_{s}}{(R-R_{s})^{2/(1+\beta)}}
\label{6.53}
\end{equation}
This is the case of repulsion and reflection from the
point $R_{s}$. The curve $R(t)$ is $\bar{C}^{2}$.
\par\medskip
\noindent{\it Matching geodesics at a white singularity:
a general solution\/}

\noindent In the vicinity of a white singularity (\ref{6.18})
we make use of equation (\ref{6.33}):
\begin{equation}
\frac{dK^{R}}{du}+\left\{ \left[ \frac{\partial _{R}
\partial _{R}r}{\partial _{R}r}+\frac{1}{2}\frac{\partial
_{R}(r_{g}/R)}{1-r_{g}/R} \right]K^{R}+\left[ 2
\frac{\partial _{t}\partial _{R}r}{\partial _{R}r} \right]
K^{0} \right\}\frac{dR}{du}=0
\label{6.54}
\end{equation}
whence, in view of $\partial _{R}r\approx 0$,
\begin{equation}
\frac{dK^{R}}{dR}+\frac{\partial _{R}\partial _{R}r}
{\partial _{R}r}K^{0}+
2\frac{\partial _{t}\partial _{R}r}{\partial _{R}r}
K^{0}\approx 0
\label{6.55}
\end{equation}

Consider a solution for which
\begin{equation}
\frac{K^{0}}{K^{R}}\approx 0
\label{6.56}
\end{equation}
so that
\begin{equation}
\frac{dt}{dR}\approx 0,\qquad \partial _{R}r\approx
(\partial _{R}\partial _{R}r)(R-R_{s})
\label{6.57}
\end{equation}
We obtain
\begin{equation}
\frac{dK^{R}}{dR}+\frac{K^{R}}{R-R_{s}}\approx 0,\quad
\frac{d}{dR}\left[ (R-R_{s})K^{R} \right]\approx 0
\label{6.58}
\end{equation}
{}from which
\begin{equation}
K^{R}\approx \frac{B}{R-R_{s}}
\label{6.59}
\end{equation}

Next we make use of
\begin{equation}
\frac{dR}{dt}=\frac{K^{R}}{K^{0}}
\label{6.60}
\end{equation}
We find
\begin{equation}
\begin{array}{l}
\omega=K^{0}=\sqrt{ m^{2}+
\displaystyle\frac{(\partial _{R}r)^{2}}
{1-r_{g}/R_{s}}(K^{R})^{2}}\approx \sqrt{m^{2}+
\displaystyle\frac{(\partial _{R}\partial _{R}r)^{2}}
{1-r_{g}/R_{s}}
\left[ (R-R_{s})K^{R} \right]^{2}}\\
\\
\qquad {}\approx\sqrt{m^{2}+
\displaystyle\frac{(\partial _{R}\partial _{R}r)^{2}B^{2}}
{1-r_{g}
/R_{s}}}={\rm const}
\end{array}
\label{6.61}
\end{equation}
Thus
\begin{equation}
(R-R_{s})dR\approx \frac{B}{\omega}dt,\quad
\frac{1}{2}(R-R_{s})^{2}\approx \frac{B}{\omega}(t-t_{s})
\label{6.62}
\end{equation}
whence it follows that
\begin{equation}
B(t-t_{s})\ge 0,\quad B(t-t_{s})=|B(t-t_{s})|=|B||(t-t_{s})|,
\quad (R-R_{s})^{2}\approx \frac{2|B|}{\omega}|t-t_{s}|
\label{6.63}
\end{equation}
so that
\begin{equation}
R-R_{s}\approx \pm\sqrt{\frac{2|B|}{\omega}}|t-t_{s}|^{1/2}
= \pm ({\rm sgn}B)[{\rm sgn}(t-t_{s})]\sqrt{\frac{2|B|}
{\omega}}|t-t_{s}|^{1/2}
\label{6.64}
\end{equation}
or
\begin{equation}
R-R_{s}\approx ({\rm sgn}B)\sqrt{\frac{2|B|}{\omega}}|t-t_{s}|^{1/2}
{\rm sgn}(t-t_{s})
\label{6.65}
\end{equation}
with ${\rm sgn}B=\pm 1$ independently of ${\rm sgn}(t-t_{s})$.
Thus we obtain
\begin{equation}
R-R_{s}\approx B\sqrt{\frac{2}{|B|\omega}}|t-t_{s}|^{1/2}
{\rm sgn}(t-t_{s})
\label{6.66}
\end{equation}
Now we have for two solutions
\begin{equation}
R-R_{s}\approx B^{[{\rm sgn}(t-t_{s})]}\sqrt{\frac{2}{\omega
|B^{[{\rm sgn}(t-t_{s})]}|}}|t-t_{s}|^{1/2}{\rm sgn}(t-t_{s})
\label{6.67}
\end{equation}
\begin{equation}
\frac{dR}{dt}\approx \frac{1}{\sqrt{2\omega}}
\frac{B^{[{\rm sgn}(t-t_{s})]}}{|B^{[{\rm sgn}(t-t_{s})]}|^{1/2}}
\frac{1}{|t-t_{s}|^{1/2}}
\label{6.68}
\end{equation}
{}From
\begin{equation}
\lim_{\delta\to 0}\frac{(dR/dt)_{t_{s}+\delta}}
{(dR/dt)_{t_{s}-\delta}}=1
\label{6.69}
\end{equation}
follows\begin{equation}
B^{[+1]}=B^{[-1]}\equiv B
\label{6.70}
\end{equation}
Thus we have finally
\begin{equation}
\begin{array}{l}
\displaystyle R-R_{s}\approx
\sqrt{\frac{2}{\omega}}\frac{B}{\sqrt{|B|}}
|t-t_{s}|^{1/2}{\rm sgn}(t-t_{s})\\
\\
\displaystyle\frac{dR}{dt}\approx
\frac{1}{\sqrt{2\omega}}\frac{B}{\sqrt{|B|}}\frac{1}
{|t-t_{s}|^{1/2}}\\
\\
\displaystyle\frac{d^{2}R}{dt^{2}}\approx
-\frac{1}{\sqrt{8\omega}}\frac{B}{\sqrt{|B|}}
\frac{{\rm sgn}(t-t_{s})}{|t-t_{s}|^{3/2}}
\label{6.71}
\end{array}
\end{equation}
We find
\begin{equation}
\frac{d^{2}R}{dt^{2}}\approx -\frac{B^{2}}{\omega^{2}}\frac
{R-R_{s}}{(R-R_{s})^{4}}
\label{6.72}
\end{equation}
This is the case of attraction by and passage through the point
$R_{s}$. The curve $R(t)$ is $\bar{C}^{1}$.
\par\medskip
\noindent{\it Matching geodesics at a white singularity:
an exceptional solution\/}

\noindent Now let us consider the case where
\begin{equation}
\partial _{R}r\approx 0, \qquad \frac{dt}{dR}\ne 0
\label{6.73}
\end{equation}
We shall be based on equations (\ref{6.34}), (\ref{6.32}):
\begin{equation}
\frac{dK^{0}}{du}+\frac{(\partial _{R}r)(\partial _{t}
\partial _{R}r)}{1-r_{g}/R}K^{R}\frac{dR}{du}=0
\label{6.74}
\end{equation}
\begin{equation}
\left( \frac{dt}{dR} \right)^{2}=\frac{\omega^{2}}{\omega^{2}
-m^{2}}\frac{(\partial _{R}r)^{2}}{1-r_{g}/R}
\label{6.75}
\end{equation}
{}From (\ref{6.75}), (\ref{6.73}) follows
\begin{equation}
\omega\approx m\ne 0
\label{6.76}
\end{equation}
We find from (\ref{6.32})
\begin{equation}
K^{R}\approx \pm\sqrt{2m(\omega-m)}\frac{\sqrt{1-r_{g}/R}}
{|\partial _{R}r|}
\label{6.77}
\end{equation}
and from (\ref{6.74})
\begin{equation}
\frac{d\omega}{dR}+({\rm sgn}K^{R})({\rm sgn}\partial _{R}r)
\frac{\partial _{t}\partial _{R}r}{\sqrt{1-r_{g}/R}}
\sqrt{2m(\omega-m)}\approx 0
\label{6.78}
\end{equation}
whence
\begin{equation}
\omega-m\approx \frac{1}{2}\frac{(\partial _{t}\partial _{R}r)
^{2}}{1-r_{g}/R}(R-R_{s})^{2},\quad \omega_{s}=m
\label{6.79}
\end{equation}
Now (\ref{6.75}) results in
\begin{equation}
\left( \frac{dt}{dR} \right)^{2}\approx \frac{(\partial _{R}r)
^{2}}{(\partial _{t}\partial _{R}r)^{2}(R-R_{s})^{2}}
\label{6.80}
\end{equation}
We have
\begin{equation}
\partial _{R}r\approx (\partial _{t}\partial _{R}r)(t-t_{s})+
(\partial _{R}\partial _{R}r)(R-R_{s})\approx
\left[ \frac{dt}{dR}+\frac{\partial _{R}\partial _{R}r}
{\partial _{t}\partial _{R}r} \right](\partial _{t}
\partial _{R}r)(R-R_{s})
\label{6.81}
\end{equation}
{}From (\ref{6.80}), (\ref{6.81}) follows
\begin{equation}
\frac{dt}{dR}\approx -\frac{1}{2}\frac
{\partial _{R}\partial _{R}r}{\partial _{t}\partial _{R}r}
\label{6.82}
\end{equation}
or, in view of (\ref{6.12}),
\begin{equation}
\left( \frac{dt}{dR} \right)_{s}=\frac{3}{4}\frac{t_{s}}{R_{s}}
\label{6.83}
\end{equation}
This gives an exceptional solution, for which there is no
problem of matching. We have from (\ref{6.81})
\begin{equation}
\left( \frac{dt}{dR} \right)_{\partial _{R}r=0}=\frac{3}{2}
\frac{t_{s}}{R_{s}}
\label{6.84}
\end{equation}
so that
\begin{equation}
\left( \frac{dt}{dR} \right)_{s}=\frac{1}{2}\left(
\frac{dt}{dR} \right)_{\partial _{R}r=0}
\label{6.85}
\end{equation}
\par\medskip
\noindent{\it Asymptotic flatness\/}

\noindent We have for $\xi\approx \xi_{n}$\begin{equation}
\chi\approx 1,\quad \frac{d\chi}{d\xi}\approx 0,\quad
r\approx R,\quad \partial _{R}r\approx 1
\label{6.86}
\end{equation}
which together with $R\gg r_{g}$ gives for the metric
\begin{equation}
ds^{2}\approx dt^{2}-[dR^{2}+R^{2}(d\theta^{2}+\sin^{2}\theta
d\phi^{2})],\qquad \frac{r_{g}^{1/2}t}{R^{3/2}}\approx \pi n,
\qquad R\gg r_{g}
\label{6.87}
\end{equation}
This holds, specifically, for
\begin{equation}
R\gg (r_{g}^{1/2}t)^{2/3},\;r_{g}
\label{6.88}
\end{equation}
\par\medskip
\noindent {\it Light cone\/}

\noindent We find for the light cone
\begin{equation}
\left| \left( \frac{dt}{dR} \right)_{\rm light} \right|=
\frac{|d\chi/d\xi|}{\sqrt{1-r_{g}/R}}\left| \frac{3}{2}\xi
-\frac{\chi}{d\chi/d\xi} \right|
\label{6.89}
\end{equation}
and for $r={\rm const}$
\begin{equation}
\left( \frac{dt}{dR} \right)_{r}=\left( \frac{R}{r_{g}} \right)
^{1/2}\left[ \frac{3}{2}\xi-\frac{\chi}{d\chi/d\xi} \right]
\label{6.90}
\end{equation}
so that
\begin{equation}
\left| \frac{(dt/dR)_{\rm light}}{(dt/dR)_{r}} \right|=
\left( \frac{1-\chi}{r/r_{g}-\chi} \right)^{1/2}
\label{6.91}
\end{equation}
this quantity
\begin{equation}
=\infty\; {\rm for}\;r=0, \quad >1\;{\rm for}\;0<r<r_{g},\quad
=1\;{\rm for}\;r=r_{g},\quad <1\;{\rm for}\;r>r_{g},\quad
=0\;{\rm for}\;r=R
\label{6.92}
\end{equation}
\par\medskip
\noindent {{\it A stationary star\/}}

\noindent {}For completeness let us touch on the case of
a stationary star. Its boundary is determined by
\begin{equation}
r(t,R)=a={\rm const},\quad R_{a}=R_{a}(t)
\label{6.93}
\end{equation}
the exterior region being
\begin{equation}
r(t,R)>a
\label{6.94}
\end{equation}
We find with the help of (\ref{6.9})
\begin{equation}
\begin{array}{l}
\eta_{a}=\eta_{a}(t),\quad \eta_{a}(0)=0,\quad R_{a}(0)=a,
\quad -\pi<\eta_{a}(t)<\pi,\quad \lim_{t\to\mp\infty}
\eta_{a}(t)=\mp\pi\\
\\
a\le R_{a}(t)<\infty,\quad
\lim_{t\to\mp\infty}R_{a}(t)=+\infty
\label{6.95}
\end{array}
\end{equation}

\section{A big crunch-bang}
\par\medskip
\noindent {\it Spacetime manifold\/}

\noindent Spacetime manifold of the standard model of the
universe is the Robertson-Walker spacetime, i.e., a product
manifold (\ref{4.1}), the three-dimensional space $S$ being
a sphere, flat Euclidean space, or a hyperboloid.
\par\medskip
\noindent{\it Synchronous coordinates\/}

\noindent Coordinates are $t$ and ''spherical'' coordinates
$r,\theta,\phi$, where
\begin{equation}
0\le r\le 1\;{\rm  for\; sphere},\quad  0\le r<\infty\;
{\rm for\; flat\; space\; and\; hyperboloid}
\label{7.1}
\end{equation}
\par\medskip
\noindent {\it Metric\/}

\noindent Metric is of the form
\begin{equation}
\begin{array}{l}
ds^{2}=dt^{2}-a^{2}(t)\left[ \frac{dr^{2}}{1-kr^{2}}+
r^{2}(d\theta^{2}+\sin^{2}\theta d\phi^{2}) \right]\\
\\
\!\!\!\!\!\!\!\!\!\!\!\!\!\!\!\!\!\!\!\!\!\!\!\!\!\!\!\!\!\!
  k=1\;{\rm for\;sphere},\quad k=0\;
{\rm for\;flat\;space},
\quad k=1\; {\rm for\;hyperboloid}
\end{array}
\label{7.2}
\end{equation}
\par\medskip
\noindent{\it Metric singularities\/}

\noindent It suffices to consider one singular hypersurface:
\begin{equation}
t=0,\qquad a(0)=0
\label{7.3}
\end{equation}
in the vicinity of which
\begin{equation}
a(t)\approx b|t|^{\beta},\quad 0<\beta<1
\label{7.4}
\end{equation}
$t<0$ and $t>0$ corresponding to the contracting and expanding
universe respectively. This is a crunch-bang singularity.
\par\medskip
\noindent{\it Matching metrics\/}

\noindent Matching metrics amounts to taking the same values
of $b,\beta$ for $t>0$ and $t<0$.
\par\medskip
\noindent{\it Geodesic equations\/}

\noindent We use relations (\ref{3.4}) through (\ref{3.6}) for
geodesics and consider ''radial'' geodesics:
\begin{equation}
(K^{i})=(K^{r},0,0),\qquad (K^{0})^{2}=h_{rr}(K^{r})^{2}+m^{2}
\label{7.5}
\end{equation}
where by (\ref{7.2})
\begin{equation}
h_{rr}=\frac{a^{2}}{1-kr^{2}}
\label{7.6}
\end{equation}
Equations (\ref{3.4}) boil down to [5]
\begin{equation}
\frac{dK^{r}}{du}+\Gamma_{rr}^{r}(K^{r})^{2}+2\Gamma_{0r}^{r}
K^{0}K^{r}=0
\label{7.7}
\end{equation}
\begin{equation}
\frac{dK^{0}}{du}+\Gamma_{rr}^{0}(K^{r})^{2}=0
\label{7.8}
\end{equation}
with
\begin{equation}
\Gamma_{rr}^{r}=\frac{kr}{1-kr^{2}},\quad \Gamma_{0r}^{r}=
\frac{\dot a}{a},\quad \Gamma_{rr}^{0}=\frac{a\dot a}{1-kr^{2}}
\label{7.9}
\end{equation}

A trivial solution is
\begin{equation}
K^{r}=0,\quad r={\rm const},\qquad \omega=K^{0}=m
\ne 0
\label{7.10}
\end{equation}
with no problem of matching.
\par\medskip
\noindent{\it Matching geodesics\/}

\noindent In the vicinity of the crunch-bang singularity
(\ref{7.3}), (\ref{7.4}) we make use of equation (\ref{7.7}):
\begin{equation}
\frac{dK^{r}}{du}+\left[\frac{kr}{1-kr^{2}}K^{r}+
2\frac{\beta}{|t|}({\rm sgn}t)K^{0}  \right]K^{r}=0
\label{7.11}
\end{equation}
{}From (\ref{7.5}), (\ref{7.6}) follows
\begin{equation}
K^{0}\ge \frac{b|t|^{\beta}}{\sqrt{1-kr^{2}}}|K^{r}|
\label{7.12}
\end{equation}
so that we obtain from (\ref{7.11})
\begin{equation}
\frac{dK^{r}}{du}+\frac{2\beta}{t}K^{r}K^{0}\approx 0
\label{7.13}
\end{equation}
or
\begin{equation}
\frac{dK^{r}}{K^{r}}+2\beta\frac{dt}{t}\approx 0
\label{7.14}
\end{equation}
whence
\begin{equation}
|K^{r}|\approx \frac{|A|}{|t|^{2\beta}},\qquad
K^{r}\approx \frac{A}{|t|^{2\beta}}
\label{7.15}
\end{equation}
Now from (\ref{7.5}) follows
\begin{equation}
(K^{0})^{2}\approx \frac{b^{2}A^{2}}{(1-kr^{2})|t|^{2\beta}}
+m^{2}\approx \frac{b^{2}A^{2}}{(1-kr^{2})|t|^{2\beta}},\qquad
K^{0}\approx \frac{b|A|}{\sqrt{1-kr^{2}}|t|^{\beta}}
\label{7.16}
\end{equation}
We have
\begin{equation}
\frac{dr}{dt}=\frac{K^{r}}{K^{0}}\approx ({\rm sgn}A)\frac
{\sqrt{1-kr^{2}}}{b}\frac{1}{|t|^{\beta}}
\label{7.17}
\end{equation}
{}from where
\begin{equation}
r-r_{0}\approx ({\rm sgn}A)\frac{\sqrt{1-kr_{0}^{2}}}{b}
\frac{1}{1-\beta}({\rm sgn}t)|t|^{1-\beta},\quad
r_{0}=r(t=0)
\label{7.18}
\end{equation}
Now we have for two solutions
\begin{equation}
(r-r_{0})^{[{\rm sgn}t]}\approx ({\rm sgn}A^{[{\rm sgn}t]})
\frac{\sqrt{1-kr_{0}^{2}}}{b(1-\beta)}({\rm sgn}t)
|t|^{1-\beta}
\label{7.19}
\end{equation}
\begin{equation}
\left( \frac{dr}{dt} \right)^{[{\rm sgn}t]}\approx
({\rm sgn}A^{[{\rm sgn}t]})\frac{\sqrt{1-kr_{0}^{2}}}{b}
\frac{1}{|t|^{\beta}}
\label{7.20}
\end{equation}
{}From the condition
\begin{equation}
\lim_{\delta\to 0}\frac{(dr/dt)_{\delta}}{(dr/dt)_{-\delta}}=1
\label{7.21}
\end{equation}
follows
\begin{equation}
{\rm sgn}A^{[+1]}={\rm sgn}A^{[-1]}\equiv {\rm sgn}A
\label{7.22}
\end{equation}
and we may put
\begin{equation}
A^{[+1]}=A^{[-1]}\equiv A
\label{7.23}
\end{equation}
Thus we have finally
\begin{equation}
\begin{array}{l}
\displaystyle r-r_{0}\approx ({\rm sgn}A)
\frac{\sqrt{1-kr_{0}^{2}}}
{b(1-\beta)}({\rm sgn}t)|t|^{1-\beta}\\
\\
\displaystyle\frac{dr}{dt}\approx ({\rm sgn}A)
\frac{\sqrt{1-kr_{0}^{2}}}{b}
\frac{1}{|t|^{\beta}}\\
\\
\displaystyle\frac{d^{2}r}{dt^{2}}\approx -({\rm sgn}A)
\frac{\beta\sqrt{1-
kr_{0}^{2}}}{b}\frac{{\rm sgn}t}{|t|^{1+\beta}}
\label{7.24}
\end{array}
\end{equation}
We find
\begin{equation}
\frac{d^{2}r}{dt^{2}}\approx -\beta(1-\beta)\left(
\frac{\sqrt{1-kr_{0}^{2}}}{1-\beta} \right)^{2/(1-\beta)}
\frac{r-r_{0}}{|r-r_{0}|^{2/(1-\beta)}}
\label{7.25}
\end{equation}
This is the case of attraction by and passage through the
point $r_{0}$.

The curve $r(t)$ is $\bar{C}^{1}$.

\section*{Acknowledgments}

I would like to thank Alex A. Lisyansky for support and
Stefan V. Mashkevich for helpful discussion.

\end{document}